\documentclass[prl,twocolumn,showpacs,amsmath,amssymb]{revtex4}
\usepackage{graphicx}
\usepackage{texdraw}

\newsavebox{\cl}
\sbox{\cl}{
\begin{texdraw}
\move (0 0) \fcir f:0.9 r:0.028 \larc r:0.028 sd:0 ed:360
\end{texdraw}}

\newsavebox{\cld}
\sbox{\cld}{
\begin{texdraw}
\move (0 0) \fcir f:1.0 r:0.028 \larc r:0.028 sd:0 ed:360
\end{texdraw}}

\newsavebox{\clu}
\sbox{\clu}{
\begin{texdraw}
\move (0 0) \fcir f:0.0 r:0.028 \larc r:0.028 sd:0 ed:360
\end{texdraw}}

\newsavebox{\ddd}
\sbox{\ddd}{
\unitlength=0.40mm%
\begin{picture}(9,5)
\put(0.3,1.2){$\diagdown$}
\put(0.5,1.4){$\diagdown$}
\put(0.7,1.6){$\diagdown$}
\put(0,-0.7){\usebox{\cld}}
\put(0,4.4){\usebox{\cld}}
\put(5.1,-0.7){\usebox{\cld}}
\end{picture}}

\newsavebox{\udd}
\sbox{\udd}{
\unitlength=0.40mm%
\begin{picture}(9,5)
\put(0.3,1.2){$\diagdown$}
\put(0.5,1.4){$\diagdown$}
\put(0.7,1.6){$\diagdown$}
\put(0,-0.7){\usebox{\clu}}
\put(0,4.4){\usebox{\cld}}
\put(5.1,-0.7){\usebox{\cld}}
\end{picture}}

\newsavebox{\dud}
\sbox{\dud}{
\unitlength=0.40mm%
\begin{picture}(9,5)
\put(0.3,1.2){$\diagdown$}
\put(0.5,1.4){$\diagdown$}
\put(0.7,1.6){$\diagdown$}
\put(0,-0.7){\usebox{\cld}}
\put(0,4.4){\usebox{\clu}}
\put(5.1,-0.7){\usebox{\cld}}
\end{picture}}

\newsavebox{\uud}
\sbox{\uud}{
\unitlength=0.40mm%
\begin{picture}(9,5)
\put(0.3,1.2){$\diagdown$}
\put(0.5,1.4){$\diagdown$}
\put(0.7,1.6){$\diagdown$}
\put(0,-0.7){\usebox{\clu}}
\put(0,4.4){\usebox{\clu}}
\put(5.1,-0.7){\usebox{\cld}}
\end{picture}}

\newsavebox{\duu}
\sbox{\duu}{
\unitlength=0.40mm%
\begin{picture}(9,5)
\put(0.3,1.2){$\diagdown$}
\put(0.5,1.4){$\diagdown$}
\put(0.7,1.6){$\diagdown$}
\put(0,-0.7){\usebox{\cld}}
\put(0,4.4){\usebox{\clu}}
\put(5.1,-0.7){\usebox{\clu}}
\end{picture}}

\newsavebox{\uuu}
\sbox{\uuu}{
\unitlength=0.40mm%
\begin{picture}(9,5)
\put(0.3,1.2){$\diagdown$}
\put(0.5,1.4){$\diagdown$}
\put(0.7,1.6){$\diagdown$}
\put(0,-0.7){\usebox{\clu}}
\put(0,4.4){\usebox{\clu}}
\put(5.1,-0.7){\usebox{\clu}}
\end{picture}}

\newsavebox{\alll}
\sbox{\alll}{
\unitlength=0.40mm%
\begin{picture}(20,10)
\put(3.2,1.2){$\diagdown$}
\put(3.4,1.4){$\diagdown$}
\put(3.6,1.6){$\diagdown$}
\put(2.9,-0.7){\usebox{\cl}}
\put(2.9,4.4){\usebox{\cl}}
\put(8.0,-0.7){\usebox{\cl}}
\end{picture}}

\newsavebox{\duuu}
\sbox{\duuu}{
\unitlength=0.40mm%
\begin{picture}(9,5)
\put(0,-0.7){\usebox{\cld}}
\put(0,4.4){\usebox{\clu}}
\put(5.1,-0.7){\usebox{\clu}}
\put(5.1,4.4){\usebox{\clu}}
\end{picture}}

\begin{document}
\title{Ground States of the Ising Model on the Shastry-Sutherland
Lattice and the Origin of the Fractional Magnetization Plateaus in
Rare-Earth Tetraborides}

\author{Yu.I. Dublenych}
\affiliation{Institute for Condensed Matter Physics, National
Academy of Sciences of Ukraine, 1 Svientsitskii Street, 79011
Lviv, Ukraine}
\date{\today}


\begin{abstract}{A complete and exact solution of the
ground-state problem for the Ising model on the Shastry-Sutherland
lattice in the applied magnetic field is found. The magnetization
plateau at the one third of the saturation value is shown to be
the only possible fractional plateau in this model. However,
stripe magnetic structures with magnetization 1/2 and $1/n$ ($n >
3$), observed in the rare-earth tetraborides RB$_4$, occur at the
boundaries of the three-dimensional regions of the ground-state
phase diagram. These structures give rise to new magnetization
plateaus if interactions of longer ranges are taken into account.
For instance, an additional third-neighbor interaction is shown to
produce a 1/2 plateau. The results obtained significantly refine
the understanding of the magnetization process in RB$_4$
compounds, especially in TmB${}_4$ and ErB${}_4$ which are strong
Ising magnets.}
\end{abstract}
\maketitle

Geometric frustrations in lattice systems result in a rich variety
of phenomena in both classical and quantum models. The
investigation of such models, and even of their ground states is,
however, a difficult problem. The first two-dimensional frustrated
quantum model whose ground states have been determined exactly was
introduced by Shastry and Sutherland in 1981 \cite{bib1}. The
Shastry-Sutherland (SS) lattice [Fig.~1(a)] is topologically
equivalent to the Archimedean lattice $3^2.4.3.4$ [Fig.~1(b)]. In
1999, the SS model was shown to describe the magnetic properties
of the compound SrCu${}_2$(BO${}_3$) \cite{bib2} synthesized in
1991 \cite{bib3}. Some time later, other quasi-two-dimensional
compounds with magnetic atoms of each layer located on a lattice
topologically equivalent to the SS one have been discovered. In
particular this concerns the rare-earth tetraborides RB${}_4$ (R =
La-Lu) \cite{bib4,bib5,bib6,bib7}. Some of these are regarded to
be classical systems, since the magnetic moments of the magnetic
ions are large. Moreover, if the crystal field effects are strong
enough, then the compounds can be described in terms of an
effective spin-1/2 SS model under strong Ising anisotropy. This is
the case of TmB${}_4$ \cite{bib4,bib5}, ErB${}_4$
\cite{bib5,bib6}, and HoB${}_4$ \cite{bib5}, where the
easy-magnetization axis is normal to SS planes.

\begin{figure}[h]
\begin{center}
\includegraphics[scale = 1.0]{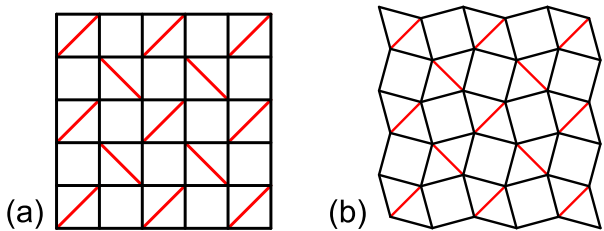}
\caption{(color online). (a) Shastry-Sutherland and (b)
Archimedean $3^2.4.3.4$ lattices. They are topologically
equivalent.}
\label{fig1}
\end{center}
\end{figure}

SS magnets exhibit fascinating and puzzling sequences of
fractional magnetization plateaus. For instance, plateaus at
$m/m_s = 1/2, 1/7, 1/8, 1/9, ...$ of saturated magnetization $m_s$
have been observed in TmB${}_4$ for temperatures below 4K when the
field was normal to the SS planes; the 1/2 plateau is the major
one. To explain the origin of this sequence of plateaus is a
challenging task. Some efforts were made to do this through the
Ising model on the SS lattice since TmB${}_4$ is a strong Ising
magnet as well as ErB${}_4$ \cite{bib5} where a single 1/2 plateau
was observed \cite{bib6}. Because of the strong frustrations, the
solution of the ground-state problem for this model is difficult
to obtain, therefore in Ref.~\cite{bib1} the Ising limit was
analyzed only for the zero field. After thirty years of research,
there is still no analytical solution for nonzero fields. There
are only some numerical results showing the existence of a single
fractional magnetization plateau at $m/m_s = 1/3$. However, the
results of numerical simulations cannot be considered as
absolutely reliable since an inappropriate finite lattice sizes
can lead to erroneous conclusions. For instance, in
Ref.~\cite{bib4} a single magnetization plateau at $m/m_s = 1/2$
was found for a system of 16 spins. More precise calculations,
however, did not confirm this conclusion \cite{bib8,bib9,bib10}.

Here, we determine a complete and exact solution of the
ground-state problem for the Ising model on the SS lattice using
our own method that has been elaborated recently
\cite{bib11,bib12}. We rigorously prove the existence of a single
1/3 plateau. As to magnetic structures with other fractional
values of $m/m_s$, they exist at the boundaries of
full-dimensional regions of the ground-state phase diagram. (A
region in the parameter space of a model is full-dimensional if
its dimension is equal to the dimension of the space). However, as
the range of interactions increases, these structures become
full-dimensional one by one and, hence, give rise to new
fractional magnetization plateaus. For example, we show that an
additional third-neighbor interaction produces a 1/2 plateau. A
similar result was obtained numerically in Ref.~\onlinecite{bib13}
but we have found three different phases with $m/m_s = 1/2$; one
of these is expected to be realized in ErB${}_4$.

Let us consider the spin-1/2 Ising model on the SS lattice in the
magnetic field
\begin{equation}
H = J_1\sum_{<ij>_1}\sigma_i\sigma_j +
J_2\sum_{<ij>_2}\sigma_i\sigma_j - h\sum_i \sigma_i.
\label{eq1}
\end{equation}
Here $\sigma_i, \sigma_j = \pm 1$ are the spin variables; $J_1$
and $J_2$ are the interaction constants (for TmB${}_4$ and
ErB${}_4$, $J_1 \approx J_2 > 0$); $<ij>_1$ and $<ij>_2$ denote
the summation over edges and diagonals [not all, see Fig.~1(a)] of
the squares, respectively; $h$ is the magnetic field.

The ground-state phase diagram for any Ising-type model is a set
of convex polyhedral cones in the parameter space \cite{bib11}. (A
polyhedral cone is the linear hull with nonnegative
coefficients---the so-called conic hull---of a set of vectors. It
is fully determined by its edges or vectors along them). This
follows from the obvious fact that the region corresponding to a
ground-state structure is a solution of a set of linear
homogeneous inequalities. The full-dimensional polyhedral cones
fill the parameter space without gaps and overlaps. Herein we
refer to a structure, which is a ground-state structure in a
full-dimensional polyhedral cone, as ``full-dimensional'' and we
refer to the corresponding edges (vectors) as the ``basic ray''
(``basic vectors'') \cite{bib11,bib12}. The convexity property
(the whole region in the parameter space, where a structure is the
ground-state one, is always convex) makes it possible to find the
ground-state structures in any point of the parameter space if all
basic rays are known as well as the ground-state structures in
them.

\begin{table}
\caption{Basic rays and basic sets of the triangle configurations
for the Ising model on the SS lattice in the magnetic field.}%
\begin{center}
\begin{tabular}{lccc}
\hline \hline
\multicolumn{1}{c}{Basic ray}&\multicolumn{1}{c}{Basic set of}&\multicolumn{1}{c}{Full-dimensional}&$\alpha$\\
$\mathbf{r}_i$ ($h$, $J_1$, $J_2$)&\multicolumn{1}{c}{configurations}& regions&\\
\hline&\\[-2mm]
$\mathbf{r}_1$ $(0, 0, -1)$&\usebox{\ddd} \usebox{\udd}
\usebox{\duu} \usebox{\uuu}
&1, $\bar 1$, 3&Arbitrary\\[2mm]
$\mathbf{r}_{2}$ $(0, -1, 2)$&\usebox{\ddd} \usebox{\dud}
\usebox{\uud} \usebox{\uuu}
&1, $\bar 1$, 2&Arbitrary\\[2mm]
$\mathbf{r}_{3}$ $(0, 1, 2)$&\usebox{\udd} \usebox{\dud}
\usebox{\uud} \usebox{\duu}
&2, 3, 4, $\bar 4$&Arbitrary\\[2mm]
$\mathbf{r}_{4}$ $(1, 0, 1)$&\usebox{\dud} \usebox{\uud}
\usebox{\duu} \usebox{\uuu}
&1, 2, 4&0\\[2mm]
$\mathbf{r}_{4}^{-}$ $(-1, 0, 1)$&\usebox{\ddd} \usebox{\udd}
\usebox{\dud} \usebox{\uud}
&$\bar 1$, 2, $\bar 4$&0\\[2mm]
$\mathbf{r}_{5}$ $(4, 1, 0)$&\usebox{\udd} \usebox{\uud}
\usebox{\duu} \usebox{\uuu}
&1, 3, 4&1/2\\[2mm]
$\mathbf{r}_{5}^{-}$ $(-4, 1, 0)$&\usebox{\ddd} \usebox{\udd}
\usebox{\dud} \usebox{\duu}
&$\bar 1$, 3, $\bar 4$&1/2\\[2mm]
\hline \hline
\end{tabular}
\end{center}
\label{table1}
\end{table}

Let us consider the rays listed in Table~I.  We shall shortly see
that these form a complete set of basic rays for the Ising model
on the SS lattice. To determine the ground-state structures in
these rays, let us rewrite the Hamiltonian (\ref{eq1}) on the SS
lattice as a single sum over all clusters in the form of a
right-angled triangle with an SS diagonal being the hypotenuse,
that is,
\begin{eqnarray}
H = \sum_{~~\usebox{\alll}}H_i =
\sum_{~~\usebox{\alll}}\left\{J_1\left(\sigma_{i0}\sigma_{i1} +
\sigma_{i0}\sigma_{i2}\right)\right.\nonumber\\
\left. + \frac{J_2}{2}\sigma_{i1}\sigma_{i2} -
h\left[\alpha\sigma_{i0} + \frac{1-\alpha}{2}\left(\sigma_{i1} +
\sigma_{i2}\right)\right]\right\}.
\label{eq2}
\end{eqnarray}
The numbers $\sigma_{i0}$, $\sigma_{i1}$, and $\sigma_{i2}$ define
a configuration (among the six unique possible ones) of the $i$-th
triangle; $\sigma_{i0}$ is the spin value in the vertex of the
right angle. With each site being a vertex for three triangles, an
arbitrary number $\alpha$ accounts for the fact that the energy
contribution of a site can be distributed among these triangles in
various ways.

We say that a structure is generated by a set of triangle
configurations if each triangle in the structure has a
configuration belonging to the set. If, in a point $(h, J_1,
J_2)$, all configurations of the set have the same energy which is
lower than the energies of all the remaining configurations, then
the structures generated by the configurations of the set are
ground-state ones in this point. Now, for each basic ray, we can
find a set of triangle configurations which generate all
ground-state structures in this ray. We call such sets ``basic
sets of configurations'' \cite{bib11}. They are given in the
second column in Table~I, and in the fourth column such a value of
the ``free'' coefficient $\alpha$ is indicated, for which all
configurations from the basic set have the same energy $H_i$ which
is lower than the energies of all the remaining configurations.
Herein solid and open circles denote spins up and down,
respectively.

Table~I represents a complete solution of the ground-state problem
for the Ising model on the SS lattice in the magnetic field. Using
this table, we first of all have to determine the full-dimensional
regions and structures. Here is an example sufficient to explain
the way for doing this: The N\'{e}el structure is generated by the
set of configurations \usebox{\udd} and
\usebox{\duu}. This set is a subset of the basic sets of
configurations for $\mathbf{r}_{1}$, $\mathbf{r}_{3}$,
$\mathbf{r}_{5}$, and $\mathbf{r}_{5}^{-}$. Hence, the N\'{e}el
structure is the ground-state one in these rays and, by virtue of
the convexity property (that is, if a structure is a ground-state
one in two points of the parameter space, then this structure is a
ground-state one in the entire line segment connecting these two
points), in the whole polyhedral cone generated by these four
vectors. A structure is full-, i.e., three-dimensional if it is
generated by a set of triangle configurations that is a subset of
at least three basic sets. All full-dimensional regions are listed
in Table~II. In its first, second, third, and fourth columns we
see, respectively: (1) the notation of the region, (2) the set of
triangle configurations generating the ground-state structures in
this region, (3) the basic rays which are edges of the region, and
(4) the conditions for the existence of the region in the plane
$(h, J_2)$. The bar over the number of a region (structure) means
that this region (structure) is symmetric to the region
(structure) with the same number but without bar with respect to
the field inversion (the flip of all spins). Now we can easily see
that the vectors listed in Table~I are basic vectors indeed and
constitute a complete set. This is so, since the full-dimensional
regions, determined from this set, fill the whole parameter space
without gaps and overlaps (see the stereoscopic diagram in Fig.~2
and the two-dimensional diagram in Fig.~3).

\begin{figure}[]
\begin{center}
\includegraphics[scale = 1.0]{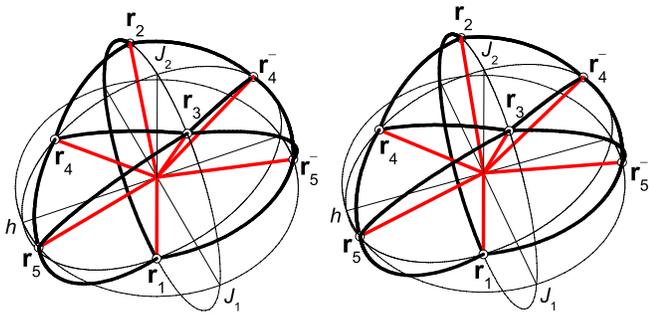}
\caption{(color online). Stereoscopic three-dimensional
ground-state phase diagram for the Ising model on the SS lattice
(see also Tables~I and II). Edges of the polyhedral cones (basic
rays) are depicted in red; faces of cones are bordered by arcs of
the unit sphere (heavy black lines). Unit circles (thin lines) in
the three orthogonal coordinate planes are figured for better
visualization. This figure is a 3-D image, therefore two parts of
the figure are almost identical.}
\label{fig2}
\end{center}
\end{figure}

\begin{table}
\caption{Basic rays and generating triangle configurations for the
full-dimensional ground-state regions of the Ising model on the SS
lattice in the magnetic field.}
\begin{center}
\begin{tabular}{cccc}
\hline \hline
Re-&Ground-state&&Existence in\\
gion&configurations&Basic rays&the plane $(h, J_2)$\\
\hline\\[-2mm]
1&~~\usebox{\uuu}&$\mathbf{r}_{1}, \mathbf{r}_{2}, \mathbf{r}_{4}, \mathbf{r}_{5}$&Always\\[2mm]
$\bar 1$&~~\usebox{\ddd}&$\mathbf{r}_{1}, \mathbf{r}_{2}, \mathbf{r}_{4}^{-},  \mathbf{r}_{5}^{-}$&Always\\[2mm]
2&~~\usebox{\dud} \usebox{\uud}&$\mathbf{r}_{2}, \mathbf{r}_{4}, \mathbf{r}_{3}, \mathbf{r}_{4}^{-}$&Always\\[2mm]
3&~~\usebox{\udd} \usebox{\duu}&$\mathbf{r}_{1}, \mathbf{r}_{5}, \mathbf{r}_{3}, \mathbf{r}_{5}^{-}$&$J_1 \geqslant 0$\\[2mm]
4&~~\usebox{\uud} \usebox{\duu}&$\mathbf{r}_{3}, \mathbf{r}_{4}, \mathbf{r}_{5}$&$J_1 \geqslant 0$\\[2mm]
$\bar 4$&~~\usebox{\dud} \usebox{\udd}&$\mathbf{r}_{3}, \mathbf{r}_{4}^{-}, \mathbf{r}_{5}^{-}$&$J_1 \geqslant 0$\\[2mm]
\hline \hline
\end{tabular}
\end{center}
\label{table2}
\end{table}

With the generating configurations for the full-dimensional
ground-state structures being known, one can easily construct the
latter. These are shown in Fig.~4, except for the fully polarized
structure 1 (see also Fig.~3 for the ground-state phase diagram).
Phase 2 is a disordered phase of Ising dimers (opposite spins on
the SS diagonals). The disorder is two-dimensional, i.e., an
extensive entropy occurs. It can be easily shown than the entropy
per site is equal to $k_B\ln 2/2$ \cite{bib1}. Phase 3 is the
N\'{e}el phase. The magnetization of the phases 1, 2, 3, and 4 is
equal to 1, 0, 0, and 1/3, respectively. Hence, there is a single
fractional plateau at $m/m_s = 1/3$, which confirms the numerical
results of Refs.~\cite{bib8,bib9,bib10}.

\begin{figure}[]
\begin{center}
\includegraphics[scale = 1.0]{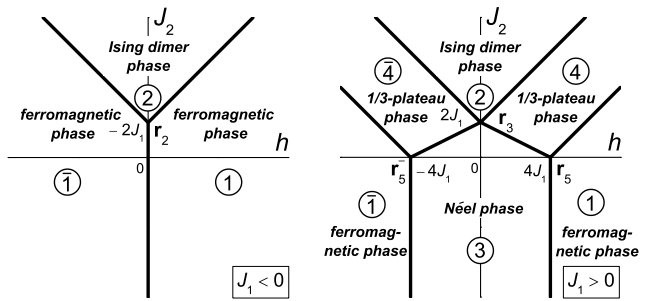}
\caption{(color online). The two-dimensional ground-state phase
diagram for the Ising model on the SS lattice (see also Tables~I
and II and Figs.~2 and 4).}
\label{fig3}
\end{center}
\end{figure}

\begin{figure}[]
\begin{center}
\includegraphics[scale = 1.0]{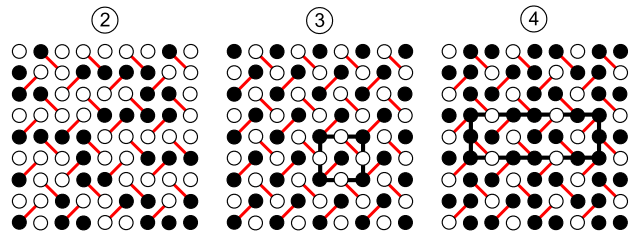}
\caption{(color online). Full-dimensional ground-state structures
for the Ising model on the SS lattice in magnetic field. Unit
cells of periodic structures are shown. Phases 2, 3, and 4 are,
respectively: the Ising dimer phase (opposite spins on the SS
diagonals), the N\'{e}el phase, and the 1/3 plateau phase. Solid
and open circles represent spins up and down, respectively.}
\label{fig4}
\end{center}
\end{figure}

\begin{table}
\begin{center}
\caption{Basic rays and triangle configurations for the
two-dimensional faces of full-dimensional ground-state regions of
the Ising model on the SS lattice in the magnetic field.}
\begin{tabular}{lccl}
\hline \hline
\multicolumn{1}{c}{Re-}&\multicolumn{1}{c}{Basic}&\multicolumn{1}{c}{Ground-state}&\multicolumn{1}{c}{Existence in}\\
gions&\multicolumn{1}{c}{rays}&\multicolumn{1}{c}{configurations}&\multicolumn{1}{c}{the plane $(h, J_2)$}\\
\hline\\[-2mm]
1, $\bar 1$&$\mathbf{r}_{1}, \mathbf{r}_{2}$&\usebox{\ddd} \usebox{\uuu}&$J_1 < 0$, $J_2 < -2J_1$\\[2mm]
1, 2&$\mathbf{r}_{2}, \mathbf{r}_{4}$&\usebox{\dud} \usebox{\uud} \usebox{\uuu}&$J_1 < 0$, $J_2 > -2J_1$\\[2mm]
1, 3&$\mathbf{r}_{1}, \mathbf{r}_{5}$&\usebox{\udd} \usebox{\duu} \usebox{\uuu}&$J_1 > 0$, $J_2 < 0$\\[2mm]
1, 4&$\mathbf{r}_{4}, \mathbf{r}_{5}$&\usebox{\uud} \usebox{\duu} \usebox{\uuu}&$J_1 > 0$, $J_2 > 0$\\[2mm]
2, 4&$\mathbf{r}_{3}, \mathbf{r}_{4}$&\usebox{\dud} \usebox{\uud} \usebox{\duu}&$J_1 > 0$, $J_2 > 2J_1$\\[2mm]
3, 4&$\mathbf{r}_{3}, \mathbf{r}_{5}$&\usebox{\udd} \usebox{\uud} \usebox{\duu}&$J_1 > 0$, $0 < J_2 < 2J_1$\\[2mm]
\hline \hline
\end{tabular}
\end{center}
\label{table3}
\end{table}

A very important advantage of our method is that it makes it
possible to determine the ground-state structures not only in
full-dimensional regions but also at their boundaries. This makes
it possible to analyze the effects of longer-range interactions.
The sets of generating configurations for the boundary structures
are given in Table~III. The sets are obtained as intersections of
corresponding basic sets. The structures observed in TmB${}_4$
emerge at the boundary between phases 3 (N\'{e}el phase) and 4
(1/3-plateau phase). They are generated by the configurations
\usebox{\udd}, \usebox{\uud}, and \usebox{\duu}. It is easy to verify that,
in addition to structures 3 and 4, these configurations generate a
sequence of stripe structures as shown in Fig.~5 and their
mixtures. The stripes contain an even number $2n$ of
antiferromagnetic chains and are bordered by the ferromagnetic
ones. We denote the structures composed of one type of stripe only
by $(3,4)_n$ ($n = 1$ for structure 4 and $n = \infty$ for
structure 3). The magnetization of structure $(3,4)_n$ is equal to
$\frac{1}{2n+1}$. The periodic mixture with the smallest unit cell
of consecutive structures $(3,4)_n$ and $(3,4)_{n+1}$ (we denote
it by $(3,4)_{n,n+1}$; see Fig.~6) has the magnetization equal to
$\frac{1}{2n+2}$. In our future paper, we will rigorously show
that structure $(3,4)_n$ ($n \geqslant 2$) can become
full-dimensional (i.e., it can produce a magnetization plateau) if
the interaction range is not less than the distance between the
successive ferromagnetic chains of the structure $(3,4)_{n-1}$. At
the boundary between regions 3 and 4, there can also occur
multidomain structures, similar to that shown in Fig.~7. They
appear if some ferromagnetic chain forms a right angle.

\begin{figure}[]
\begin{center}
\includegraphics[scale = 1.0]{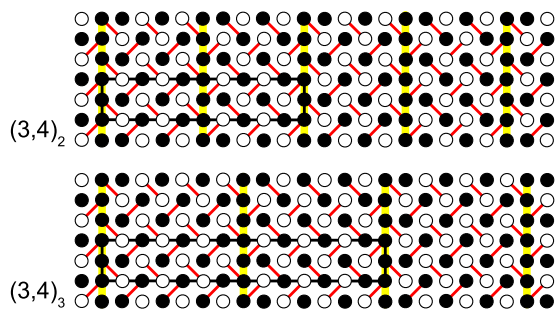}
\caption{(color online). Ground-state structures $(3,4)_2$ and
$(3,4)_3$ ($m/m_s$ = 1/5 and 1/7) at the boundary between phases 3
and 4. The structures consist of ferro- (on yellow background) and
antiferromagnetic chains. Unit cells are indicated.}
\label{fig5}
\end{center}
\end{figure}

\begin{figure}[]
\begin{center}
\includegraphics[scale = 1.0]{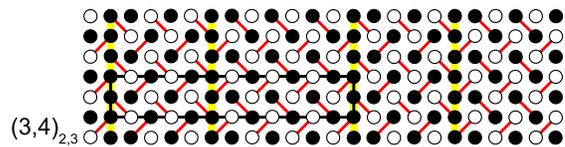}
\caption{(color online). Ground-state structure $(3,4)_{2,3}$
($m/m_s$ = 1/6) at the boundary between regions 3 and 4. This
structure is the periodic mixture with the smallest unit cell of
structure $(3,4)_2$ and $(3,4)_3$ (see Fig.~5).}
\label{fig6}
\end{center}
\end{figure}

Let us consider some other boundaries. The structures at the
boundary between regions 1 and 3 (see Table III and Fig.~8) are
constructed in the following way: Squares with SS diagonals should
be chosen in such a manner that any couple of squares should have
no common sites. At these diagonals, the spins are orientated
downwards and all the remaining spins are orientated upwards. The
structures at the boundary between regions 1 and 4 (Fig.~8) are
determined by a single condition: there cannot be two neighboring
spins downwards neither along SS diagonals nor along the edges. If
an additional antiferromagnetic (as in TmB${}_4$) third-neighbor
interaction $J_3$ along diagonals of ``empty'' (i.e., without SS
diagonals) squares is included, then, from all possible structures
at the boundary between regions 1 and 3 and 1 and 4, only those
structures become full-dimensional in which all the ``empty''
squares have the configuration \usebox{\duuu} (Fig. 8), since the
energy should be as small as possible. These structures constitute
disordered phases with magnetization 1/2. One can show that the
width of corresponding plateaus is equal to $4J_3$. Thus, contrary
to the conclusion of Ref.~\cite{bib13}, we can say that additional
third-neighbor interaction is sufficient for the stabilization of
a 1/2 plateau. If $J_3 < 0$, then among all structures at the
boundary between phases 1 and 4 only the ordered structure
$(1,4)_{-}$ shown in Fig.~8 becomes full-dimensional. It produces
a 1/2 plateau as well. We suppose that this structure emerges in
ErB${}_4$ since $J_3$ is expected to be negative in this compound
\cite{bib5}. Interaction $J_3$ lifts also the degeneracy of the
Ising dimer phase. If $J_3 > 0$ ($J_3 < 0$), then among all the
structures of this phase, the structure with the minimum energy in
the one for which spins at $J_3$ bonds are different (identical).
These structures are shown in Fig.~9. The Ising dimer phase exists
just at the boundary between these.

\begin{figure}[h]
\begin{center}
\includegraphics[scale = 1.0]{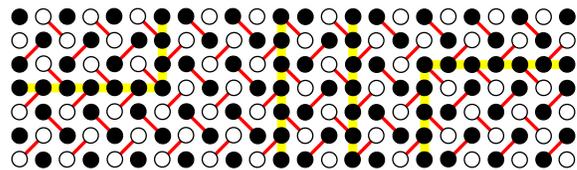}
\caption{(color online). A multidomain structure at the boundary
between regions 3 and 4 ($m/m_s = 0$). Left and right
ferromagnetic chains form right angles.}
\label{fig7}
\end{center}
\end{figure}

\begin{figure}[h]
\begin{center}
\includegraphics[scale = 1.0]{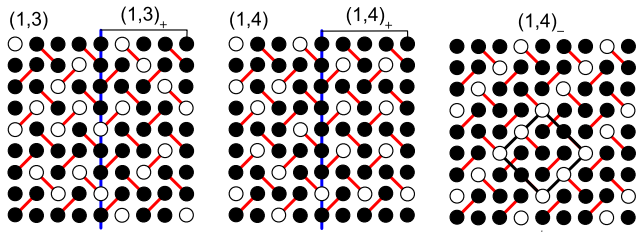}
\caption{(color online). Ground-state structures at the boundaries
between regions 1 and 3 and 1 and 4 (see also Table III). First
and second figures present disordered structures. The right-hand
parts of these figures show the arrangements of spins (still
disordered) if an additional third-neighbor interaction $J_3 > 0$
is included. All the "empty" squares have then the
configuration \usebox{\duuu}. The third figure shows an ordered
structure that becomes full-dimensional if $J_3 < 0$. All the
structures $(1,3)_{+}$, $(1,4)_{+}$, and $(1,4)_{-}$ produce a 1/2
plateau. (See the text for details).}
\label{fig8}
\end{center}
\end{figure}

\begin{figure}[ht]
\begin{center}
\includegraphics[scale = 1.0]{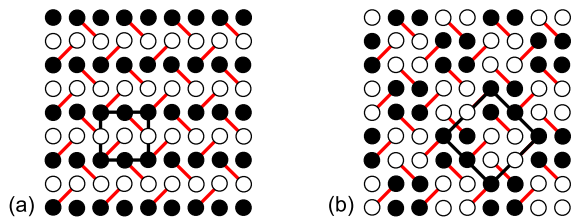}
\caption{(color online). Ground-state structures of the Ising
model in the magnetic field on the SS lattice with additional
third-neighbor interaction: (a) $J_3 > 0$ (collinear phase), (b)
$J_3 < 0$ (chessboard phase). The Ising dimer phase exists at the
boundary between these two phases.}
\label{fig9}
\end{center}
\end{figure}

Some ground-state structures, obtained here analytically, coincide
with those found numerically by different authors. Structure 4 was
determined in Refs.~\cite{bib9,bib10,bib13,bib14}, structure 3
(N\'{e}el phase) in Refs.~\cite{bib4,bib9,bib10}, structures 2
(Ising dimers) in Refs.~\cite{bib1,bib10}. It seems that
structures with magnetization 1/4, 1/5, and 1/6 coincide with
those determined numerically in Ref.~\cite{bib14} for a
spin-electron model (we cannot be sure, since the SS bonds are not
indicated there). This is quite clear because the spin-electron
interaction lifts the degeneracy at the boundary between regions 3
and 4 and some structures become full-dimensional. Our structures
with magnetization $\frac{1}{2n+1}$ are different from those shown
in Ref.~\cite{bib4}: all antiferromagnetic chains are shifted. The
same concerns the structure with magnetization 1/3 shown in
Ref.~\cite{bib8}.

The main conclusion of our study is that the Ising model with
additional long-range interactions is sufficient to explain the
origin of fractional magnetization plateaus in rear-earth
tetraborides which are strong Ising magnets (TmB${}_4$ and
ErB${}_4$). The long-range interactions are
Ruderman-Kittel-Kasuya-Yosida ones \cite{bib4,bib13}, since
rear-earth tetraborides are good metals. Here we have studied the
role of the first- and second-neighbor interactions and
rigourously proved that they produce a single fractional plateau
at 1/3 of saturated magnetization. We have also partially analyzed
the effect of the third-neighbor interaction and thus have shown
that it produces at least five new phases; for three of these the
magnetization is equal to 1/2. One of these 1/2-plateau phases is
expected to emerge in ErB${}_4$. Having analyzed the structures at
the boundary between the N\'{e}el phase and the 1/3-plateau phase,
we have also found the stripe structures with $m/m_s = 1/n~ (n
\geqslant 4)$; some of these have been observed in TmB${}_4$. To
finally explain the emergence of the sequence of magnetization
plateaus in TmB${}_4$, the effect of further-neighbor interactions
should be investigated.

We are grateful to T. Verkholyak for drawing our attention to the
problem of fractional magnetization plateaus in SS compounds and
for useful discussions.

\end{document}